\documentclass[global,final]{svjour}

\def\union{{\cup}}

\usepackage{alltt}
\usepackage{epsfig}
\usepackage{a4}
\usepackage{times}
\usepackage{amsfonts}

\def\lrarrow{\leftrightarrow \kern-8pt \rightarrow}
\def\imply{\Rightarrow}
\def\frightarrow{\rightarrow \kern-11pt /~~}

\def\ctrust{{\rm ~Trusts~}}
\def\CAND{\sc AND}
\def\COR{\sc OR}
\def\CXOR{\sc XOR}
\def\beq{\begin{eqnarray}}
\def\eeq{\end{eqnarray}}
\def\2{\frac{1}{2}}
\newtheorem{defin}{Definition}
\newtheorem{proposal}{Proposal}

\title{Local and Global Trust Based on the Concept of Promises}

\author{Jan Bergstra\inst{1} \and Mark Burgess\inst{2}}

\institute{Informatics Institute, Science park 904, 1098 XH, the Netherlands, Email: j.a.bergstra@uva.nl\and Department of Computing, Oslo University College, Norway, Email: mark.burgess@iu.hio.no}
\date{20th September 2006, minor edits 23 December 2009}

\begin{document}
\newcommand{\trust}[1]{\stackrel{\tau:#1}{\longrightarrow}}
\newcommand{\promise}[1]{\stackrel{\pi:#1}{\longrightarrow}}
\newcommand{\policy}{\stackrel{P}{\equiv}}
\newcommand{\field}[1]{\mathbb{#1}}

\maketitle

\begin{abstract}
We use the notion of a promise to define {\em local trust} between
agents possessing autonomous decision-making. An agent is trustworthy if it
is expected that it will keep a promise.  This definition satisfies
most commonplace meanings of trust.  Reputation is then an estimation
of this expectation value that is passed on from agent to agent.

Our definition distinguishes {\em types} of trust, for different
behaviours, and decouples the concept of agent reliability from the
behaviour on which the judgement is based. We show, however, that
trust is fundamentally heuristic, as it provides insufficient
information for agents to make a rational judgement.  A global
trustworthiness, or {\em community trust} can be defined by a
proportional, self-consistent voting process, as a weighted
eigenvector-centrality function of the promise theoretical graph.
\end{abstract}


\section{Introduction}

\begin{quote}
{\em I don't trust him. We're friends.}\\ ~~~~~~ --Bertolt Brecht
\end{quote}

The decision to trust someone is a policy decision. Although the
decision can be made {\em ad hoc}, our common understanding of trust
is that it is based on a gathering of experience, i.e. a process of
learning about the behaviour and reputation of someone in a variety of
scenarios.  Our particular policy might weight certain sources and
behaviours more heavily than others and no one can tell us what is the
right thing to do.  Hence trust is intimately connected with personal
autonomy.

In this paper, we define trust in the spirit of this personal
autonomy, by basing it directly on the concept of how reliably a
promise is kept. A promise is also an autonomously made declaration of
behaviour, that is highly individual, moreover it carries with it
the notion of a theme (what the promise is about)\cite{promiseidea}. By combining
promises with reliability, we thus have a natural definition of trust
that satisfies well-understood rules for revising both the logical
aspects of policy and the statistical observations made about agents'
behaviours. We show that this viewpoint satisfies the desirable properties
for use in computer security schemes. 
Note that our aim in this work is not to design a technology for
building trust, but rather to analyse the precepts for what trust actually
means to users.

\begin{figure}[ht]
\begin{center}
\psfig{file=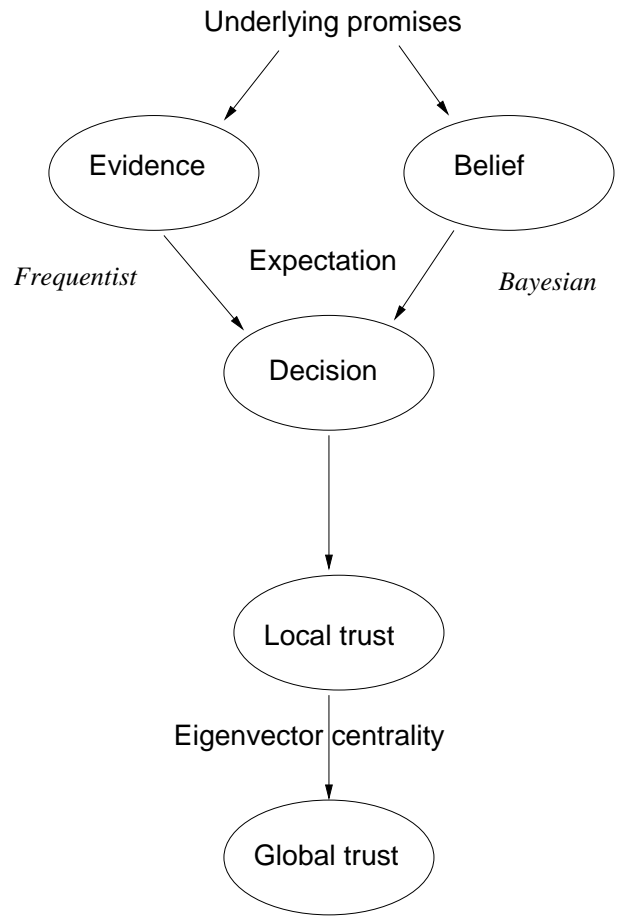,width=8cm}
\caption{The chain of trust from verifiable promises to local trust
by an agent, to global or community trust which we interpret as reputation.\label{trust}}
\end{center}
\end{figure}

The plan for this paper is as follows (see fig. \ref{trust}), We discuss the
notion of trust from a pragmatic and philosophical point of view in
order to settle on what properties trust should have. We show that
common expressions of trust are often ambiguous, but that we can
resolve this ambiguity by defining agent trust as the expectation of
keeping a given promise.  Using the graphical notions of promises, we
can then establish a notion of global trust in certain cases.

\subsection{Trust}

The concept of trust is both well known and widely used in all kinds
of human interactions. Trust is something that humans hold both for
one another or sometimes for inanimate objects (``I trust my computer
to give the right answer'').  In computer systems, the concept of
trust is especially used in connection with security. In risk analysis
one considers a secure system to be one in which every possible risk
has either been eliminated or accepted as a matter of policy. Trust is
therefore linked to the concept of policy in a fundamental way.

Trust is also discussed in the case of network security protocols, for instance, 
in the case where keys are exchanged. The classic dilemma of key distribution
is that there is often a high level of uncertainty in knowing the
true originator of a secure identifier (cryptographic key). One therefore
hopes for the best and, beyond a certain threshold of evidence ``trusts'' the
assumption of ownership. Several protocols claim to manage such trust
issues, but what does this really mean?

In spite of the reverence in which the concept is held, there is no
widely accepted technical definition of trust. This has long be a
hindrance to the discussion and understanding of the concept. The
Wikepedia defines: ``Trust is the belief in the good character of one
party, they are believed to seek to fulfil policies, ethical codes,
law and their previous promises.''  In this paper, we would like to
address the deficiencies of discussions of trust by 
introducing a meta-model for understanding trust. Our model can be
used to explain and describe common trust models like ``trusted third
parties'' and the ``web of trust''.

\subsection{Promises -- autonomous claims}

Trust is an evaluation that can only be made by an individual. No one
can force someone to trust someone else in a given situation. This
basic fact tells us something important about how trust should be defined.

Recently, one of us has introduced a description of autonomous behaviour in
which individual agents are entirely responsible for their own
decisions\cite{burgessDSOM2005,siri1,siri2,siri3}.  Promise theory is
a graphical model of policy.  The basic responsibility of an
agent to be true to its own assertions is an important step towards a
way of describing trust.

Promise theory is useful in this regard because all agents are
automatically responsible for their own behaviour and only their own
behaviour. Responsibility is not automatically transitive
between autonomous agents: it has to be arranged through explicit
agreement between agents in a controlled way; hence one avoids
problems such as hidden responsibility that make the question of
whether to trust an individual agent complex.

In this paper, we argue that the concept of trust can be defined
straightforwardly as a {\em valuation} of a promise -- specifically the {\em
expectation} of autonomous behaviour. When we say that we trust
something, we are directing this towards the instigator of some
promise, whether implicit or explicit.  Moreover {\em reputation} is
simply what happens to trust as it is communicated about a network,
i.e. it is a `rumour' that spreads epidemically throughout a network along
different paths, and hence develops into a path-dependent estimate of
trustworthiness.

The matter of evidence-gathering, in order to justify the expectation
value of keeping a promise is subtle, and so we shall discuss this in
some detail. We argue that there is insufficient information in the
notions of trust or reputation to make a reliable estimate of
trustworthiness. Thus trust is an inherently ambiguous concept; each
valuation of trustworthiness is, in essence, an essentially {\em ad
hoc} policy.

\section{Prior work}

There is an extensive literature on trust in computer
science\cite{lapadula1,mcilroy1,winkler2,patton04technologies,sang-can,huynh2004a}. Much
of it is concerned with generating protocols for the purpose of
determining the validity of public keys and other identity tokens, or
criticizing these mechanistic views in a wider security perspective.
Here we are mainly concerned with general ideas about trust and
reputation.

We find the recent work of Kl\"uwer and Waaler to be of interest from
the viewpoint of logic\cite{klwer05trustworthiness,relativetrust}.
These authors present a natural reasoning system about trust which includes
the notion of {\em ordering} by levels of trustworthiness.

The work that seems closest to ours may be found in ref.
\cite{beth1} and ref. \cite{jossang1}. 
Here the authors distinguish between trust and reputation and provide
an epidemic-like procedure for valuating the trust based on some
inference rules and numerical measures that are essentially
reliabilities. The calculation is hence mainly appropriate for a
frequentist interpretation of probability. The authors in ref.
\cite{beth1} are unable to
distinguish trust about different issues, or relate these in their
model. In ref.  \cite{jossang1}, an attempt is made at motivating
trust types but the underlying properties of these types is not
completely clear.

In our proposal:
\begin{enumerate}
\item We allow for multiple sources (types) for which trust and reputation are valuated.

\item Our combinatorics are based on logic and on Bayesian probability estimates, 
which are more appropriate
estimators for the small amounts of experience involved.
\end{enumerate}

Other work which we find valuable includes social viewpoints of trust
(see ref. \cite{trust1} for a review).  This work brings in the matter
of human value judgements, which we feel is an important issue in any
definition of trust, since it is humans who make the final decisions
in practice.  From a sociological viewpoint, there are many forms of
currency on which to build trust. Some of these are based on the
outcomes of stand-offs such as economic games, bargaining situations
and so on\cite{axelrod2}.  Promises have already been shown to incorporate
these considerations neatly within their framework\cite{siri2}.


\section{Common usage of trust and reputation}

As with most words, the English word `trust' has a number of related
meanings which are worth documenting for reference and comparison.

\begin{itemize}
\item Trust implies a confidence or faith character:
e.g. one  ``trusts in friends and family''.

\item It might be based on an assessment of reliability: e.g. ``A trustworthy employee''

\item A related, but not identical meaning has to do with presumed safety.
It also means to permit something without fear. ``I trust the user to
access the system without stealing.'' Such trust can be betrayed.

This is different because the feeling of safety is not a rationally determined
quantity, whereas reliability is observable and measurable.  Thus
there is both a rational and an irrational aspect to trust.

\item A final meaning of trust is the expression of hope, i.e. and
expectation or wish: "I trust you will behave better from now on";

Trust is therefore about the suspension of disbelief. It involves a
feeling of benevolence, or competence on the part of the trustee.

Trust of this kind expresses an acceptance of risk, e.g.  a jewelry
store trusts that passers-by will not smash a plate glass window very
often to steal displayed goods, but rather trusts that the windows
will improve sales. There could therefore be an economic decision
involved in risk-taking.
\end{itemize}

Reputation is a related notion to trust. We understand this to mean a
received judgement, i.e. an evaluation of an agent's reliability based
on hearsay. Reputation spreads like an epidemic process, but it is
potentially modified on each transmission. Thus, from a given source,
several reputations might emerge by following different pathways
(histories) through a network.


\section{A typed definition of trust}

An agent that is known to keep its promises is considered trustworthy
by any normal definition of trust i.e. the agent would be reliable and
predictable such that one could put aside one's doubts about whether
it might fail to live up to its assertions.

It seems natural then to associate trust with one agent's expectation
of the performance of another agent in implementing its promises.
This could seem like an unnecessarily narrow definition, but it turns
out to be more general than one might expect. What about trust in
matters that have not yet occurred?  Clearly, trust could be
formulated about a future {\em potential promise}.  i.e. a promise
does not have been made for us to evaluate its likely reliability.  The
usefulness of promises is that they encapsulate the relevant
information to categorise intentions and actions.

\begin{proposal}[Trust]
Trust can be defined as an {\em agent's expectation} that a promise will
be kept. It is thus a probability lying between 0 and 1.
\end{proposal}

We shall define ``an agent's expectation'' in detail below, and we
shall additionally give meaning to the concepts of when an agent is
deemed to be {\em trustworthy} or {\em trusting} which are global
concepts, different from merely {\em trusted}. This proposal has a
number of positive qualities. To begin with it separates the {\em
experiential} aspect of trust from the {\em nature of the actions} on
which it is based. Thus in terms of philosophy of science, it makes a
clean distinction between empirical knowledge (expectation) and
theoretical knowledge (a promise).

Our definition is specific.  The concept of trust, as normally applied in
computer science is rather universal and non-specific: either one
trusts another agent or one does not; however, it is seldom that we
trust or distrust anyone or anything so completely. Our definition is a
{\em typed} definition, i.e. we gauge trust separately for each
individual kind of promise -- and this is where promises provide a convenient
notation and conceptual stepping stone. We assume that promises are a
more fundamental notion than trust.

According to our definition, trust is a reliability rating made by some
agent that is able to observe two agents involved in a promise. We
hesitate to call this a reliability {\em measure}: for reasons that we
shall make clear, there is normally insufficient evidence on which to
base a proper reliability estimate, in the sense of reliability
theory\cite{hoyland1}.

A reputation is little more than a rumour that spreads epidemically
throughout a network. Common ideas about reputation include.
\begin{itemize}
\item ``A general opinion of someone.''
\item ``A measure of someone's standing in the community.''
\end{itemize}
Reputation is not necessarily related to trustworthiness.  One could
have a reputation based on how much money an agent spends, or how much
fuel it uses.  What characterizes a reputation, as opposed to a
personal observation or evaluation, is that it is passed on. One does
not observe the characteristic first hand.

\begin{proposal}[Reputation]
Reputation can be defined as a valuation of some agent's past or
expected behaviour that is communicated to another agent.
\end{proposal}

We clarify and develop these basic proposals in the remainder of the paper.
In particular trust will be revisited in more
detail in section 8.


\subsection{Promises}

To base our notion of trust on promises, we review the basic concepts
from promise theory. Promises are closely linked to the idea of
policy, or declarations of autonomous decision-making. Indeed, we define
policy to be simply a set of promises, since one can always phrase
decisions about how to respond to future events and scenarios as promises
about what one will do.

Consider a general set of $N$ agents $A_i$, where $i=1,\ldots,N$.  We
denote agents by capital Roman letters, and shall often use nicknames
$S$ for promise-sender or giver, $R$ for promise receiver and $T$ for
third parties, to assist the discussion.

\begin{defin}[Promise]
A promise is an autonomous specification of unobserved or intended
behaviour. It involves a {\em promiser} and one or more promisees to
whom the promise is directed. The {\em scope} of the promise is the
set of agents who have knowledge of the promise that has been made.
Each promise contains a promise body $b$ that describes the content of
the promise. We denote a promise from agent $S$ to an agent $R$, with
body $b$ by:
\beq
S \promise{b} R
\eeq
\end{defin}
The body $b$ of every promise contains a {\em type} $t(b)$ and often
an additional constraint. Promise types distinguish the qualitative
differences between promises, and additional constraint attributes
distinguish the quantitative differences. For each promise body $b$,
there is another promise body $\neg b$ which represents the negation
of $b$. We shall assume that $t(\neg b)=t(b)$ and that $\neg\neg b=b$.
The negation of $b$ refers to the deliberate act of not performing
$b$, or what ever is the complementary action of type $t(b)$.

Promises fall into two basic complementary kinds, which we can think of
as promises for giving and taking, or {\em service} and {\em usage}.
A promised exchange of some service $s$ requires one of each kind:
\beq
A_1 \promise{s} A_2 ~~~{\rm or}~~~ A_1 \promise{+s} A_2\nonumber\\
A_2 \promise{U(s)} A_1 ~~~{\rm or}~~~ A_2 \promise{-s} A_1
\eeq
Exclusive promises are those which cannot physically be realized at
the same time.  This does not mean that incompatible promises cannot
be made, it means that they are meaningless and could lead to problems
for the agent.

\begin{defin}[Incompatible promises $\#$]
When two promises originating from an agent are incompatible, 
they cannot be realized physically at the same time. We write 
\beq
A_1 \promise{ b_1} A_2 ~\#~
A_1 \promise{ b_2} A_3
\eeq
If $A_2 = A_3$, we may omit the agents and write $b_1 \# b_2$.
\end{defin}
It would probably be unwise for an agent to trust another agent that
made simultaneous, incompatible promises. Of course this is a policy
decision for each individual agent to make.

Breaking a promise is not the same as not keeping a promise. It is
an explicit contradiction. Again, confidence in an agent's promise-keeping
ability is reduced when it makes contradictory promises.

\begin{defin}[Promise conflict]
Two or more promises to an agent are in conflict if at least one promise is
contradicted by another.  We define a conflict as the promising
of exclusive promises
\beq
A_1 \promise{ b_1} A_2, ~ A_1 \promise{ b_2} A_2~ {\rm with}~  b_1 \# b_2
\eeq
\end{defin}
Clearly promising $b$ and $\neg b$ would be excluded.

\subsection{Composition of parallel promises}

We can compose trivial {\em bundles} of promises between a single pair
of agents by union. Using proof notation:
\beq
\frac{a \promise{b_1} b,a \promise{b_2} b,\ldots a \promise{b_N} b,}
{a \promise{b_1 \union b_2 \union \ldots b_N} b}
\eeq
where the non-overlap of independent type regions is not necessarily
assumed, but helps to make sense of this (the definition should still
work even if the types overlap).  The composition of promises in a
serial fashion is non-trivial and only has meaning in a minority of
cases, where autonomy is relinquished.

A promise made conditionally on a Boolean condition $C$, known to the
promising agent is written:
\begin{defin}[Conditional promise]
A conditional promise is written:
\beq
A_1 \promise{b|C} A_2
\eeq
i.e. $A_1$ promises $b$ to $A_2$ if the condition $C$ is true.
\end{defin}

Note that a condition can also be the subject of a promise.
We write
\beq
A_1 \promise{{\rm T}(C)} A_2
\eeq
for the promise from $A_1$ to $A_2$ to ensure that the condition $C$
holds.  Now the combination of a conditional promise and the promise
of the condition holding leads to an unconditional promise as follows:
\beq
\frac{
A_1 \promise{b|C} A_2, ~
A_1 \promise{{\rm T}(C)} A_2
}
{
A_1 \promise{b} A_2
}
\eeq
A promise is not truly a promise unless the truth of the condition is
also promised.

\subsection{General notation for promises}

The following notation for promises has been designed to be clear and
pragmatic, avoiding potential recursion difficulties of promises about
promises. We begin with the kind of basic promise from one agent to
another and then generalize this:
\begin{enumerate}  

\item 
The preferred form of a promise (first kind) is written. 
\beq
S \promise{b} R
\eeq
This is a local and autonomously made promise. This is
equivalent to the more general notation:
\beq
S[S] \promise{b} R[R].
\eeq
i.e. $S$ promises $b$ to $R$.

\item A promise of the second kind allows obligation:
\beq
S[T] \promise{b} R
\eeq
i.e. $S$ promises $R$ that it will oblige $T$ to act as if it had
promised $b$ to $R$. If $T$ is autonomous, this is forbidden and has
no influence on $T$.

\item A promise of the third kind allows indirection.
\beq
S \promise{b}R[T]
\eeq
i.e. $S$ promises to $R$ that $S$ will do $b$ for $T$.

\item The most general form of a promise:
\beq
S[T] \promise{b} D[U]
\eeq
i.e. $S$ promises $D$ that $b$ will act as if it had promised $b$ to $U$.
If $T$ is autonomous, this is forbidden and has no
influence of $T$.
\end{enumerate}
We have potentially a need for all of these variants.

\begin{example}
Promises about policy in which one does not inform
the promise recipient (e.g. the Border Gateway Protocol (BGP) policy)
may be written:
\beq
S \promise{b} S[R]
\eeq
i.e. $S$ makes a promise only to itself to honour $b$ toward $R$ (e.g.
suppose $b$ is a promise to use packet-data received).  This is the case, for
instance, in the processing of Access Control Lists by most network
devices: the sender of data has no {\em a priori} idea of whether the
device will accept it.
\end{example}

Most of the promises we shall consider in the definition of trust will
be of the form
\beq
A_2 \promise{b} A_1[A_3]
\eeq
i.e. a neighbouring agent promises us that it will do something
for some third party (where the third party might also be us).

\subsection{A general expression for trust}

Trust is somehow complementary to the idea of a service promise. This
is suggested by the intuition that a promise to {\em use} a service
implies a measure of trust on the part of the receiver.  We consider
trust a directed relationship from a {\em truster} to a {\em
trustee}. Moreover, it is a judgement or {\em valuation} of a promise
performed entirely by the {\em truster}.

We need a notation to represent this, similar to that for promises. In
the spirit of the promise notation, we write the general case as:
\beq
S[T] \trust{b} R[U]
\eeq
meaning that $S$ trusts $R$ to ensure that $T$ keeps a promise
of $b$ to $U$.

In most cases, this is too much generality. In a world of autonomous
agents, no agent would expect agent $S$ to be able to ensure anything
about agent $T$'s behaviour. The more common case is therefore with
only three parties
\beq
A_1[A_2]  \trust{b}    A_2[A_3]
\eeq
i.e. agent $A_1$ trusts agent $A_2$ to keep its promise towards some
third-party agent $A_3$.  Indeed, in most cases $A_3$ might also be
identified with $A_1$:
\beq
A_1[A_2]  \trust{b}    A_2[A_1]
\eeq
which, in turn, can be simplified to
\beq
A_1 \trust{b} A_2.
\eeq
In this case, trust is seen to be a dual concept to that of a promise.
If we use the notation of ref. \cite{siri2}, then we can write trust
as one possible valuation $v: \pi \rightarrow [0,1]$ by $A_1$ of the
promise made by $A_2$ to it:
\beq
A_1[A_2]  \trust{b}    A_2[A_1]~ \leftrightarrow ~v_1(A_2 \promise{b} A_1)
\eeq
This is then a valuation on a par with economic valuations of how much
a promise is worth to an agent\cite{siri2}. The recipient of a promise
can only make such a valuation if it knows that the promise has been
made.
\begin{proposal}
Trust of an agent $S$ by another agent $R$ can exist if agent $R$ is
informed that agent $S$ has made a promise to it in the past, or if
the recipient of the promise $R$
is able to infer by indirect means that $S$ has made such a
promise.
\end{proposal}
Thus any agent can formulate a trust policy towards any other agent.
The only remaining question is, on what basis should such a judgement
be made?

Our contention is that the most natural valuation to attach to trust is
an agent's estimate of the expectation value that the promise will be
kept, i.e. an estimate of the reliability of the agent's
promise. 
\beq
A_1[A_2]  \trust{b}    A_2[A_1]~ \policy ~E_1(A_2 \promise{b} A_1)
\eeq
where $\policy$ means `is defined by policy as', and the expectation
value $E_R(\cdot)$, for agent $R$ has yet to be defined (see Appendix
A for these details).  We note the essential difficulty: that such
valuations of reliability are not unique. They are, in fact, entirely
subjective and cannot be evaluated without ad hoc choices of a number
of free parameters. We return to this point below.


\section{Cases: The underlying promises for trust idioms}

To ensure that our definition of trust is both intuitive and general,
we present a number of `use-cases' below and use these to reveal, in
each case, the expectation of a promise that underlies the trust.
In each case, we write the declarations of trust, in notation,
in words, and as an expectation value of an underlying promise.
In some cases, the expressions of trust are ambiguous and support several
interpretations which can only be resolved by going to a deeper explanation
in terms of promises.
\begin{itemize}

\item {\em I trust my computer to give the right answer.}
This could literally mean that one trusts the computer, as a potentially unreliable piece
of hardware:
\beq
{\rm Me} \trust{\rm answer}{\rm Computer} \policy E_{\rm {\rm Me}}({\rm Computer} \promise{\rm answer} {\rm Me})
\eeq
i.e. I expect that the computer will keep its (implicit) promise to furnish me with the correct answer.

However, there is another interpretation.
We might actually (even subconsciously) mean that we trust the company that produces
the software (the vendor) to make the computer deliver the right answer when asked, i.e.
I expect the promise by the vendor to me, to make the computer give me the right answer, will
be kept.
\beq
[{\rm Me}][{\rm Computer}]\trust{\rm answer} [{\rm Vendor}][{\rm Me}]~~~~~~~~~~~~~~~~~~~~~~~~~~~~~~~~~~~~\nonumber\\
\policy E_{\rm Me}\left( [{\rm Vendor}][{\rm Computer}] \promise{{\rm Answer}}  [{\rm Me}][{\rm Me}]\right)
\eeq
In either case, the relationship between the promise, the expectation and the trust is the same.

\item {\em I trust the identity of a person (e.g. by presence, public key or signature).}

This is one of the classic problems of security systems, and we find that the simple
statement hides a muddle of possibilities. It has many possible interpretations; however, in
each case we obtain clarity by expressing these in terms of promises.

\beq {\rm Me} \trust{\rm Authentic}{{\rm Signature}} \policy E_{{\rm Me}}({\rm Signature}
\promise{\rm Authentic} {\rm Me})
\eeq
In this version, we place trust in the implicit promise that a credential makes of being
an authentic mark of identity. This is a simple statement, but we can be sceptical of the
ability of a signature to make any kind of promise.

\beq {\rm Me}[{\rm Signature}] \trust{\rm Authentic}{{\rm Certifier}}[\rm Me]~~~~~~~~~~~~~~~~~~~~~~~~~~~~~~~~~~~~ \nonumber\\
\policy E_{{\rm Me}}({\rm Certifier}[{\rm Signature}]
\promise{\rm Authentic} {\rm Me})
\eeq
i.e. I trust a Certifying Agency to ensure that the implicit promise
made by the credential to represent someone is kept. Or I expect the
certifying agency (possibly the originator of the signature himself)
to keep a promise to me to ensure that the signature's promise to me
is kept (e.g. the technology is tamper-proof).

Yet a third interpretation is that the trust of the key is based
on the promise to verify its authenticity, on demand. This is the
common understanding of the ``trusted third party''.
\beq
{\rm Me} \trust{\rm verify~ key} {\rm Certifier}
\policy E_{\rm Me}\left(
{\rm Certifier} \promise{\rm verify~key} {\rm Me}
\right)
\eeq
i.e. I trust that the key has been authorized and is verifiable by the named
Certification Agency. This last case avoids the problem of why one should
trust the Certifying Agency, since it refers only to the verification service
itself.

\item A similar problem is encountered with currency denominations, e.g.
pound notes, dollars, or Euros. These tokens are clearly not valuable
in and of themselves; rather they represent value. Indeed, on British
Pound notes, the words ``I promise to pay the bearer on demand the sum of ... X
pounds'' is still found, with the printed signature of the Chief
Cashier. Indeed, the treasury will, if pressed, redeem the value of
these paper notes in gold. Thus trust in a ten pound note may be
expressed in a number of ways.

We trust the note to be legal tender: i.e.
\beq
{\rm Me} \trust{\rm legal} {\rm Note} \policy E_{\rm Me}
\left(
{\rm Cashier} \promise{\rm gold|note} {\rm Me}
\right)
\eeq
we expect that the chief cashier will remunerate us in gold on presenting
the note. Alternatively, we assume that others will promise to accept the
note as money in the United Kingdom (UK):
\beq
{\rm Me} \trust{\rm legal} {\rm Note} \policy E_{\rm Me}
\left(
{\rm S} \promise{\rm U({\rm note})} {\rm Me}
\right),~~ S \in UK
\eeq
Interestingly neither dollars nor Euros make any much promise. Rather, the
dollar bill merely claims ``In God we trust''.

\item {\em Trust in family and friends.}

This case is interesting, since it is so unspecific that it could be
assigned almost any meaning. Indeed, each agent is free to define its
meaning autonomously. For some set of one or more promises ${\cal P}^*$,
\beq 
{\rm Me} \trust{\rm {\cal P}^*}{\{\rm Family}\} \policy E_{\rm {\rm Me}}\left( \bigcup_{i\in *} \{{\rm Family}\} \promise{\rm
{\cal P}_i} A_i\right)
\eeq
i.e. for some arbitrary set of promises, we form an expectation about the likelihood
that family and friends would keep their respective promises to the respective
promisees. These promises might, in fact, be hypothetical and the evaluations mere
beliefs. On the other hand, we might possess actual knowledge of these transactions,
and base judgement on the word of one of these family/friend members to keep their
promises to the third parties:
\beq 
{\rm Me} \trust{\rm {\cal P}^*}{\{\rm Family\}} \policy E_{\rm {\rm Me}}\left( \bigcup_{i\in *} {\{\rm Family\}} \promise{\rm
{\cal P}_i}{\rm Me} [A_i]\right)
\eeq

\item {\em A trustworthy employee.}

In this case, one bases trustworthiness is based more on a history of delivering on promises
made in the context of work, e.g.:
\beq
{\rm Boss} \trust{\rm Deliver} {\rm Employee} \policy E_{\rm Boss}({\rm Employee} \promise{\rm Deliver} {\rm Boss})
\eeq

\item {\em I trust the user to access the system without stealing.}

Here the promise is not to steal. The promise does not have to have
been made explicitly. Indeed, in civil society this is codified into
law, and hence all agents implicitly promise this by participating in
that society.

\item {\em ``I trust you will behave better from now on!''}

This can be understood in two ways. In the first interpretation, this
is not so much an evaluation of trust as it is a challenge (or even
warning) to the agent to do better. Alternatively, it can be taken literally
as an expression of belief that the agent really will do better. In the latter
case, it is:
\beq
{\rm Me} \trust{\rm Do~ better} {\rm You} \policy E_{\rm Me}\left(
{\rm You} \promise{\rm Do~better} {\rm Me}
\right)
\eeq

\end{itemize}


\section{Expectations of ensembles and compositions of promises}

We are not done with policy's intrusion into the definition of
expectation.  Since promises can be composed according to
straightforward rules, we must be able to compute two distinct things:
\begin{enumerate}
\item The expectation of a composition of promises that coexist.
\item The composition of expectations from different ensembles.
\end{enumerate}
The difference between these is analogous to the difference between
the combinations of experimental data into ensembles for computing
probabilities, and the composition of different probable inputs in
fault trees (with $\CAND$, $\COR$, $\CXOR$, etc).

We have already discussed the composition of data sets into ensembles,
the effect this has on probabilities, and how this is expressed in terms
of the basic expectation values in section \ref{ensemble}

We shall have need to define the meaning of the following in order to
determine the trust deriving from compound promises:

\begin{enumerate}
\item The expectation of incompatible promises.
\item The expectation of a composition of parallel promises between a pair of agents.
\item The expectation of a composition of serial promises between a chain of agents.
\end{enumerate}

\subsection{Parallel promise (bundle) expectation}

When promises are made in parallel, the question arises as to how much
to trust them as a bundle. Should one ever base one's trust on a
complete package or bundle of promises?  This is a subjective
judgement based on whether certain promises are related in the view of
the promisee.  If one makes an expectation valuation for each promise
individually, does it make sense to combine them as probabilities,
e.g. in the manner of a fault tree\cite{burgessbook2,hoyland1}.  One
is used to the probability composition rules for binary logic of
independent events.

\begin{itemize}
\item ($\CAND$): If the promisee is
dependent on several mutually reinforcing promises, then $\CAND$
semantics are a reasonable assumption. In a security situation, this
might be reasonable.  The multiplicative combination rule means that each
additional promise that must be in place reduces the total trust that the
promiser will keep all of its promises proportionally.

\item ($\COR$) Here one says that if one or more promises are kept, then
trustworthiness is reinforced. This is an optimistic policy which
seems to suggest that the promisee is understanding about the promiser's
potential difficulties in keeping a promise. This cannot be applied
to incompatible promises.

\item ($\CXOR$): An alternative scenario is to have a number of promises that are
alternatives for one another. For instance, mutually exclusive
conditional promises that behave like a switch: e.g.
\beq
S &\promise{x|y} &R\nonumber\\
S &\promise{x'|\neg y} &R,
\eeq
i.e. $S$ promises $x$ to $R$, iff $y$, else it promises $x'$.

\item ({\sc RANKED}) If the promises are ranked in their importance to the recipient,
then the measure of trust associated with the package is best judged by weighting the
importance appropriately. Referring to the discussion in section \ref{ensemble}, this
admits a general convex combination of contributions for ranking an $\COR$ (see below).
\end{itemize}
Let us consider how these are represented as functions.

\begin{defin}[Expectation of a promise bundle]
Let $S$ (sender) and $R$ (recipient) be agents that make a number of promises in parallel,
the composition of a bundle of parallel promises $S \promise{b^*} R$
is a function $F_R$ of the expectations of the individual promises:
\beq
E_{R}\left(S \promise{b^*} R\right) \policy F_{R} \left( E_{R}\left( S \promise{b_1} R\right),E_{R}\left( S \promise{b_2} R\right),\ldots\right)
\eeq
\end{defin}

The function $F_R$ is a mapping from $N$ promise expectations to a new expectation value:
\beq
F_R : [0,1]^N  \rightarrow [0,1]
\eeq
Several such functions are known from reliability theory, e.g. in fault tree
analysis (see for instance ref. \cite{hoyland1}). Examples include,
\beq
F^{\rm AND}_{R}
\left(S \promise{b^*} R\right) 
&=& 
\prod_i 
E_{R}\left(S \promise{b_i} R\right)\\\nonumber\\
F^{\rm OR}_{R}
\left(S \promise{b^*} R\right) 
&=& 
1-\prod_i 
\left( 1 - E_{R}\left(S \promise{b_i} R\right)\right)\nonumber\\
&\simeq& \sum_i E_{R}\left(S \promise{b_i} R\right) ~\pm~ O(E^2)\\
F^{\rm XOR}_{R}
\left(S \promise{b^*} R\right) 
&\simeq& 
1-\prod_i 
\left( 1 - E_{R}\left(S \promise{b_i} R\right)\right)\nonumber\\
&\simeq& \sum_i E_{R}\left(S \promise{b_i} R\right) ~\pm~ O(E^2).
\eeq
where $O(E^2)$ denotes terms or the order of the probability squared, which are small.
A further possibility is to take a weighted mean of the promise
estimates.  This better supports the view in section \ref{ensemble}
about different sizes ensembles and their relative weights. There
might be additional (irrational) reasons for giving priority to
certain promises, e.g. leniency with respect to a difficult promise.

To combine the different possibilities (analogously to fault trees) one could
first reduce products of $\CAND$ promises into sub-bundles, then recombine these
using a weighted estimate.
\beq
F^{\sc RANKED}_{R} &\policy& \sum_i \alpha_i E_{R}\left(S \promise{b_i} R\right)\nonumber\\
\sum_i \alpha_i &=& 1
\eeq

Note that, due to the reasoning of probability theory, the expectation
of something AND something else is less than the probability of
either.  This might be seen as pessimistic as far as trust is
concerned. We have to make a policy decision about whether or not to
place any weight on the combined expectation of a bundle of promises,
or whether to decide to only allow individual expectations.

For example, suppose an agent makes two contradictory promises
about services levels, e.g. promise to respond in 4ms and promise
to respond in 5ms.
\beq
S &\promise{4}& R\nonumber\\
S &\promise{5}& R
\eeq
Formally, this is a conflict, since both promises cannot be true
at the same time.  The trust in each individual promise can be
estimated independently for the two promises.  The agent reliability
expectations of delivering ``4'' or ``5'' units of service are:
\beq
R \trust{4} S = E_R(4) &=& p(4) = 0.1\\
R \trust{5} S = E_R(5) &=& p(5) = 0.2
\eeq
Then we can consider what the expectation of the combination of promises
is. If the agent $S$ makes both promises simultaneously, the
expectation of the combined promises will be:
\beq
E_R(4 ~\CXOR~ 5) &\simeq& \frac{(e_4\, E_R(4) + e_5\, E_R(5))}{(e_4+e_5)}
\eeq
where $e_4$ is our estimate of likelihood the agent can deliver ``4''
and $e_5$ is the estimate of likelihood of delivering ``5''.
These beliefs can be based on many potential sources of information, chosen as a matter
of policy; one possibility is to simply identify $e_4 \policy E_R(4)$
and $e_5 \policy E_R(5)$. Thus a simple policy solution could be
to take 
\beq
E_R(4 ~\COR~ 5)~ \policy~ \frac{E_R(4)^2+E_R(5)^5}{E_R(4)+E_R(5)} = 0.17
\eeq
i.e. in general a sum of squares.

\subsection{Incompatible promise expectation}

For incompatible promises we must have at least complementary behaviour ({\sc NOT}):
\beq
E_A(S \promise{\neg b} R) &=& 1 - E_A(S \promise{b} R)\nonumber\\
F_R(E_R(S \promise{\neg b} R)) &=& 1 - F_R(E_R(S \promise{b} R))
\eeq
Ideally incompatible promises would not be made, without conditionals
to select only one of the alternatives. 

In the case of $\CAND$ it is necessary already to resolve the 
ambiguity in the meaning of the combination of incompatible
promises. It is by definition a logical impossibility for incompatible
promises to be kept. Thus, while we cannot prevent an agent from promising
such nonsense, our expectation of the combination ought 
to be zero.
\begin{defin}[Expectation of incompatible promises with $\CAND$]

The expectation of incompatible promises,
\beq
F_R\left(A_1 \promise{ b_1} A_2 ~\CAND ~A_1 \promise{ b_2} A_2\right) \equiv 0 ~~{\rm when}~  b_1 \# b_2
\eeq
is defined to be zero for any rational agent.
\end{defin}
Hence, in the example above,
\beq
E_R(4 ~\CAND ~5) &=& 0.
\eeq

\subsection{Serial promise expectation and transitivity of trust}

Several systems base their operation on the idea that trust is to some
extent transitive.  ``The Web of Trust'' notion in public key
management idea proposes that trust can be conferred transitively. This
is not a property of promises, so it is of interest to consider how
this works. In other words, if $A_1$ trusts $A_2$ to do $b$, and $A_2$
trusts $A_3$ to do $b$, then $A_1$ will often trust $A_3$ to do $b$. Here
$b$ is generally taken to be ``reveal one's true identity''.  This
notion does not fit well with a promise theory interpretation of trust
because it is type-unspecific.

This is easy to see by noting that
\beq
A_1 \promise{b} A_2 , A_2 \promise{b} A_3 \not\imply A_1 \promise{b} A_3
\eeq
i.e. if $A_1$ makes a promise of $b$ to $A_2$ and $A_2$ makes the same
promise to $A_3$, it does not follow that $A_1$ has made any promise to
$A_3$.

An unspecific trust model might conform to the following property:
\beq
(i)~~ (A_1 \ctrust A_2) , (A_2 \ctrust A_3) \imply A_1 \ctrust A_3
\eeq
In terms of promises, we would interpret this to mean that, if $A_1$
trusts $A_2$ (to keep promises to $A_1$) and $A_2$ trusts $A_3$ (to keep
promises to $A_2$) then $A_1$ should trust $A_3$ to keep promises to
$A_1$. This is far from being a rational policy, since there is no
evidence passed on about the reliability of agents.
A less problematic alternative is:
\beq
(ii)~~ (A_1 \trust{\rm inform} A_2) , (A_2 \trust{b} A_3) \imply A_1[A_3] \trust{b} A_3[A_2]
\eeq
If $A_1$ trusts $A_2$ (to inform it about its relations with $A_3$)
and $A_2$ trusts $A_3$ (to keep its promise of $b$
to $A_2$), then $A_1$ trusts that $A_3$ is trustworthy in its promise of $b$ to $A_2$.

The matter of serial promises is one of diverging complication.
We make some brief notes about the problems associated with serial
promises, and leave the potentially extensive details for elsewhere.
The problems with trusting a distributed collection of promises
are
\begin{enumerate}
\item Promises are not common knowledge, so we do not have all the information.
\item Promises are not transitive.
\end{enumerate}

\begin{figure}[ht]
\begin{center}
\psfig{file=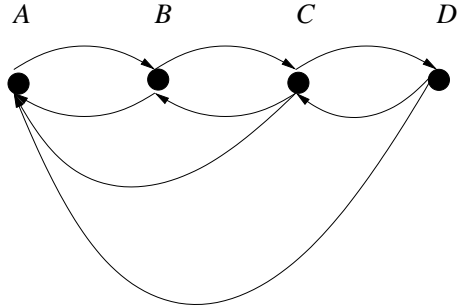,width=6cm}
\caption{A chain of trust promises to transfer some valuation of trust in 
one direction (only), from node
$a$ to each agent up to node $d$. This method is unreliable because
nodes $b$ and $c$ are under no obligation to pass on the correct
value. Note that these are promise arrows, not trust arrows.\label{chain}}
\end{center}
This is clearly a fragile and somewhat complicated structure.  An
alternative approach is to avoid chains of greater length than one,
and also eliminate the extraneous and essentially impotent promises
from the chain, as in fig. \ref{mr}. However, this leads us merely
back to the notion of a centralization, either in the form of a
trusted party for all agents, or as a complete peer-to-peer graph.
\end{figure}

Knowledge about the promises and the local evaluations by the agents
can only be guaranteed by making chains of promises between the agents
to share this knowledge.
\beq
A_1 &  \promise{\rm tell\,rep}~ A_2 ~\promise{\rm tell\,rep}& A_3\nonumber\\
A_1 & \stackrel{\pi:U({\rm tell\,rep})}{\longleftarrow}~ A_2 ~\stackrel{\pi:U({\rm tell\,rep})}{\longleftarrow}& A_3
\eeq
In order to pass on the necessary information about trust to a third
party, it must be relayed. Expectation of a chain of promises depends
on a chain of such trust and Use(trust) promises. However, each agent
in the chain agrees only to trust the previous agent. There is no
automatic agreement to trust the previous members. If one were to make
an explicit promise to trust each agent's information about trust,
this would require a promise graph like the one in fig. \ref{chain}.
In order to remove the ambiguity of the trust promises, we must use a
different {\em promise type} for trust about each agent in the graph.
i.e. the trust passed on from agent $a$ must retain this label in
being transferred. However, here one has a paradox: if an agent is potentially
unreliable, then it can easily lie about this information.
Such serial chains are, in general fraught with uncertainty, thus 
agents might well choose, as a matter of policy, to disregard
reputations.
\begin{figure}[ht]
\begin{center}
\psfig{file=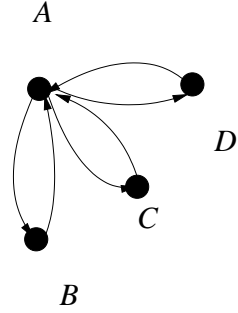,width=3cm}
\caption{A more reliable approach of passing on the trust node $a$ holds
on to nodes $b$, $c$ and $d$.\label{mr}}
\end{center}
\end{figure}


\section{Reputation}

We have defined a reputation to be simply a valuation of something
(not necessarily a promise) received by an agent about some other
agent. A natural basis for reputation (and one that is used on
`reputation systems' in computing) is the valuation of
trustworthiness.  Here we consider the effect that such transmission
of information has on the local trust within a network of agents.

\subsection{Borrowed trust}

Suppose that and agent $T$ trusts an agent $S$ to keep its promise to
$R$ with probability $E_T\left( S\promise{b} R\right)$, and suppose
that this agent $T$ promises to transmit this as $S$'s reputation to
another agent $U$.  $U$'s estimate of the trustworthiness of $T$'s
communication is
\beq
U \trust{\rm reputation} T \policy E_U\left( T \promise{\rm reputation} U\right)
\eeq
Can we say what $U$'s expectation for the reliability of the original
promise $a\promise{b} c$ should be? In spite of the fact that
probabilities for independent events combine by multiplication, it
would be presumptuous to claim that
\beq
E_U\left(S\promise{b} R\right) = E_U\left( T \promise{\rm reputation} U\right)
E_T\left( S\promise{b} R \right),
\eeq
since $U$ does not have any direct knowledge of $E_T\left(
S\promise{b} R \right)$, he must evaluate the trustworthiness and
reliability of the source.

Suppose we denote the communicated value of  $E_T\left( S\promise{b} R \right)$
by  ${\cal E}_{U\leftarrow T}\left( S\promise{b} R \right)$, then one could conceivably
(and as a matter of rational policy) choose to define
\beq
E_U\left(S\promise{b} R\right) \policy E_U\left( T \promise{\rm reputation} U\right)
{\cal E}_{U\leftarrow T}\left( S\promise{b} R \right).
\eeq
With this notation, we can conceivably follow historical paths through
a network of promises.

However, it is important to see that no agent is obliged to make such
a policy.  Thus trust and reputation do not propagate in a faithfully
recursive manner.  There is, moreover, in the absence of complete
and accurate common knowledge by all agents, an impossibility of eliminating the
unknowns in defining the expectation values.

\subsection{Promised trust}

Trust is an evaluation that is private to an agent.  This evaluation
can be passed on in the form of a communication (leading to
reputation), or it can be passed on as a promise to trust.

\begin{itemize}
\item $S$ promises $R$ that $S$ will trust $R$: $S \promise{\tau=0.6} R$.
\item $S$ promises $R$ that $S$ will trust $T$: $S \promise{\tau=0.6} R[T]$.
\end{itemize}
Why would anyone promise a party ($R$) to trust $T$ without telling $R$?
One reason is that there might be strategic bargaining advantages to doing this\cite{schelling1}.

\subsection{Updating trust with reputation}

An agent can use the reputation of another agent as a sample of
evidence by which to judge its trustworthiness. It can then attach a
certain weight to this information according to its belief, in order
to update its own trust.  The weighted addition modifies the old trust
value $T$ with the new reputation data $R$.
\beq
E \mapsto \frac{w_{\rm new} R + w_{\rm old} T}{w_{\rm new}+ w_{\rm old}}
\eeq
This is indistinguishable from a Bayesian update.



\section{Global Measures of Trust}\label{central}

Which are the most trusted agents in a network?
Trust has so far been measured at the location of each individual
agent. The valuation is private.  A trust valuation becomes an agent's
reputation when the valuation is passed on to others.  The passing-on
includes a revisional belief process too; this is also a Bayesian
posterior probability update process, just like the case of basing
trust on different ensembles in section \ref{ensemble}.

Let us postulate the existence of a vector of received trusts that is
available to any particular agent. The agent is then able to combine
this information to work out a global measure, which we can call
{\em community trust}. This is analogous to the graphical security model in
\cite{burgessC12}.

The trust matrix $T$ is defined as follows. The $(A,B)$-th element of the
matrix
\beq
T_{AB}(b) \equiv E_A(B \promise{b} *)
\eeq
is $A$'s trust in $B$ with respect to all promises of type $b$.

\begin{defin}[Community trust (Trustworthiness and trustingness)]
The global or community trust is defined by the principal eigenvectors
of $T$ and $T^{\rm T}$. Since this is a transmitted quantity by definition
it is a reputation. 

The global reputations for being {\em trustworthy} $\vec W$ are
defined by the normalized components of the principal eigenvector of
the transpose matrix:
\beq
T_{BA} W_B = \lambda  W_A.
\eeq

The global reputations for being {\em most trusting} $\vec S$ are
defined by the normalized components of the principal eigenvector
\beq
T_{AB} S_B = \lambda  S_A.
\eeq
\end{defin}
An agent is said to be trusting if it assigns a high probability of
keeping its promises to those agents that it trusts. An agent is said
to be trustworthy if other agents assign it a high probability of
keeping promises to it.

Observe that, in the absence of labels about specific agent
relationships, the concepts of {\em trustworthiness} and {\em
trustingness} for an agent $A$ are properties of the global trust graph that
has $A$ as a source, and not of an individual agent, since they are derived
from relationships and by voting.

We can easily show that this has the property of a proportional vote.
Let $v_i$ denote a vector for the trust ranking, or connectedness of
the trust graph, of each node $i$. Then, the trustworthiness of node $i$ is
proportional to the sum of the votes from all of $i$'s nearest
neighbours, weighted according to their trustworthiness (i.e. it is just
the sum of their trust valuations):
\beq
v_i \propto\sum_{j={\rm neighbours\ of\ }i} v_j \ \ .
\label{evc1}
\eeq
This may be more compactly written as 
\beq
v_i = ({\rm const}) \times \sum_j T_{ij} v_j \ ,
\label{evc2}
\eeq
where $T$ is the {\em trust graph adjacency matrix}, whose entries
$T_{ij}$ are 1 if $i$ is a neighbour of $j$, and 0 otherwise.  We can
rewrite eqn. (\ref{evc2}) as
\beq
{T}\,\vec{v} =\lambda \vec v \ .
\label{evcfin}
\eeq 
Now one sees that the vector is actually an eigenvector of the trust
matrix $T$. If $T$ is an $N\times N$ matrix, it has $N$ eigenvectors
(one for each node in the network), and correspondingly many
eigenvalues. The eigenvalue of interest is the principal eigenvector,
i.e. that with highest eigenvalue, since this is the only one that
results from summing all of the possible pathways with a positive
sign.  The components of the principal eigenvector rank how
self-consistently `central' a node is in the graph.  Note that only
ratios $v_i/v_j$ of the components are meaningfully determined. This
is because the lengths $|\vec v|= \sqrt{\sum_i v_iv_i}$ of the eigenvectors are not determined
by the eigenvector equation. We normalize them here by setting the
highest component to 1.  This form of well-connectedness is termed
'eigenvector centrality' \cite{bonacich1} in the field of social
network analysis, where several other definitions of centrality
exist.

\begin{figure}[ht]
\begin{center}
\psfig{file=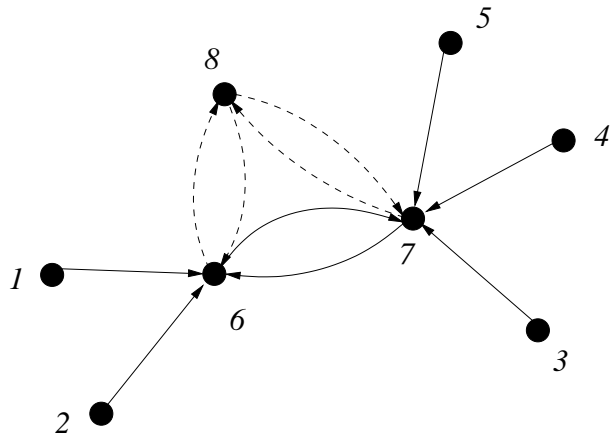,width=8cm}
\caption{An example trust graph. For simplicity all trust arrows are
assumed of the same type, e.g. trust in the promise to pay bills.
Dashed lines are lines which will be removed in the second example.\label{exb}}
\end{center}
\end{figure}

Note this does not assume any transitivity of trust, it says simply:
each agent's trust worthiness is equal the sum of all the other
agents' trust measures (as if they are voting), weighted so that the
most trustworthy agents' opinions are weighted proportionally highest.
It is a proportional representation vote by the agents about one another.

\subsection{Example of global trust}

Consider a number of promises of a single type, e.g.  agents promise
to pay their bills in various service interactions. Each payee then
rates its expectation of the payer and makes this information globally
available as a public measure of its local trust. Referring to
fig. \ref{exb}, we assume the following local trusts:
\beq
1& \trust{\rm pay}& 6 = 0.2\\\nonumber
2& \trust{\rm pay}& 6 = 0.3\\\nonumber
3& \trust{\rm pay}& 7 = 0.1\\\nonumber
4& \trust{\rm pay}& 7 = 0.1\\\nonumber
5& \trust{\rm pay}& 7 = 0.1\\\nonumber
6& \trust{\rm pay}& 7 = 0.6\\\nonumber
7& \trust{\rm pay}& 6 = 0.5\\\nonumber
6& \trust{\rm pay}& 8 = 0.8\\\nonumber
8& \trust{\rm pay}& 6 = 0.2\\\nonumber
7& \trust{\rm pay}& 8 = 0.8\\\nonumber
8& \trust{\rm pay}& 7 = 0.3
\eeq
The trust matrix is thus
\beq
T = \left(
\begin{array}{ccccccc|c}
0.0 & 0.0 & 0.0 & 0.0 & 0.0 & 0.2 & 0.0 & 0.0\\
0.0 & 0.0 & 0.0 & 0.0 & 0.0 & 0.3 & 0.0 & 0.0\\
0.0 & 0.0 & 0.0 & 0.0 & 0.0 & 0.0 & 0.1 & 0.0\\
0.0 & 0.0 & 0.0 & 0.0 & 0.0 & 0.0 & 0.1 & 0.0\\
0.0 & 0.0 & 0.0 & 0.0 & 0.0 & 0.0 & 0.1 & 0.0\\
0.0 & 0.0 & 0.0 & 0.0 & 0.0 & 0.0 & 0.6 & 0.8\\
0.0 & 0.0 & 0.0 & 0.0 & 0.0 & 0.5 & 0.0 & 0.8\\\hline
0.0 & 0.0 & 0.0 & 0.0 & 0.0 & 0.2 & 0.3 & 0.0\\
\end{array}
\right)
\eeq
Note that the bars delineate the dashed lines which will be removed in the
second example.
The normalized right eigenvector $\vec S_8$ represents how trusting
the agents are. The left eigenvector $\vec W_8$ (or the eigenvector of
the transpose matrix) represents the global trustworthiness:
\beq
\vec S_8 = \left(
\begin{array}{c}
0.21\\
0.31\\
0.10\\
0.10\\
0.10\\
1.00\\
0.94\\
0.50\\
\end{array}
\right), ~~~
\vec W_8 = \left(
\begin{array}{c}
0\\ 
0\\
0\\
0\\
0\\
0.55\\
0.65\\
1.00\\
\end{array}
\right)
\eeq
Thus, agent 8 is the most trustworthy. Agents 1 to 5 are not trustworthy at
all in this scenario, since we have not rated any promises made by
them. Agent 6 is the most trusting of all, since it gives a large
amount of trust to agent 8.  Thus, these two agents colour the global
picture of trust significantly through their behaviours.

We note that the agents with zero trust ratings are all recipients of
promises; they do not make any promises of their own. These are
suppliers of whatever service or good is being sold; they do not
promise payments to anyone, hence no one needs to trust them to pay
their bills. The reader might find this artificial: these agents might
make it their policy to trust the agents even though they have made no
promise. In this case, we must ask whether the trust would be of the
same type or not: i.e. would the buyers trust the suppliers to pay
their bills, or would their trust be based on a different
promise, e.g. the promise to provide quality goods.

By contrast, the agents who are not trusted are somewhat trusting by
virtue of receiving such promises of payment.

Suppose we eliminate agent number 8 (by removing the dashed lines in
the figure), let us see how the ranking changes when we delete this
important agent. Now agent 6 still remains the most trusting, but
agent 7 becomes the most trusted, once again mainly due to agent 6's
contribution.

\beq
\vec S_7 = \left(
\begin{array}{c}
0.37\\
0.55\\
0.17\\
0.17\\
0.17\\
1.00\\
0.92\\
\end{array}
\right), ~~~
\vec W_7 = \left(
\begin{array}{c}
0\\
0\\
0\\
0\\
0\\
0.91\\
1.00\\
\end{array}
\right)
\eeq
We can note that the symmetries of the graph are represented in the 
eigenvector in a natural way.

\subsection{Boundaries and allegiances}

Canright and Monsen have defined regions of a graph, based on the
structures that arise naturally from eigenvector
centrality\cite{roles}. This has been further developed for directed
graphs in ref. \cite{burgessroles}. Trust is sometimes associated with
maintaining certain boundaries or allegiances.  The global trust model
proposed above falls into a natural landscape based on the graph, that
is characterized by local maxima. Agents cluster naturally into
distinct hills of mutual trust, separated by valleys of more tenuous
trust, in the centrality function.

This characterization is a useful way of identifying a community
structure.  Humans are not very good at understanding boundaries: they
understand identities. e.g. a company name, but where is the real
boundary of the company or computer system? Its tendrils of influence
might be farther or closer than one imagines. The topology of
underlying promises offers a quantifiable answer to this question.
Such allegiances can be compared to the notion of a coalition in game
theory\cite{morgenstern1,rapoport1}.


\section{Trust architectures}

Trust is closely associated with information dissemination. There are
essentially only two distinct models for achieving information
distribution: centralization and {\em ad hoc} epidemic flooding.
Alternatively one might call them, central-server versus peer-to-peer.

Two so-called trust models are used in contemporary technologies
today, reflecting these approaches: the Trusted Third Party model
(e.g. X.509 certificates, TLS, or Kerberos) and the Web of Trust (as
made famous by the Pretty Good Privacy (PGP) system due to Phil
Zimmerman and its subsequent clones). Let us consider how these models are
represented in terms of our promise model.

\subsection{Trusted Third Parties}

\begin{figure}[ht]
\begin{center}
\psfig{file=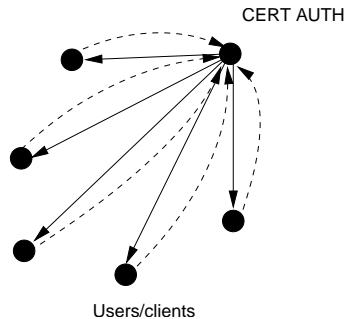,width=4.5cm}
\caption{\small The Trusted Third Party, e.g. TLS or Kerberos. A
special agent is appointed in the network as the custodian of
identity. All other agents are expected to trust this.
The special agent promises to verify the authenticity of an
object that is shared by the agents. In return for this service, the
agents pay the special agent.\label{thirdparty}}
\end{center}
\end{figure}

The centralized solution to ``trust management'' is the certificate
authority model, introduced as part of the X.509 standard
and modified for a variety of other systems (See
fig. \ref{thirdparty})\cite{itut1,x509,rfc3280}. In this model, a central authority has the
final word on identity confirmation and often acts as a broker between
parties, verifying identities for both sides.

The authority promises (often implicitly) to all agents the legitimacy
of each agent's identity (hopefully implying that it verifies this
somehow).  Moreover, for each consultation the authority promises that
it will truthfully verify an identity credential (public key) that is
presented to it.  The clients and users of this service promise that
they will use this confirmation. Thus, in the basic interaction, the
promises being made here are:
\beq
{\rm Authority} &\promise{\rm Legitimate}  &{\rm User}\\
{\rm Authority} &\promise{\rm Verification} & {\rm User}\\
{\rm User} &\promise{U({\rm Verification})} &{\rm Authority}
\eeq
To make sense of trust, we look for expectations of the promises
being kept. 
\begin{enumerate}
\item The users expect that the authority is legitimate, hence they
trust its promise of legitimacy.
\item The users expect that the authority verifies identity correctly, hence
they trust its promise of verification and therefore use it.
\end{enumerate}
Users do not necessarily have to be registered themselves with
the authority in order to use its services, so it is not strictly
necessary for the authority to trust the user. However, in registering
as a client a user also promises its correct identity, and the authority
promises to use this.
\beq
{\rm User} &\promise{\rm Identity}& {\rm Authority}\\
{\rm Authority} &\promise{U({\rm Identity})}&  {\rm User}
\eeq
One can always discuss the evidence by which users would trust the
authority (or third party).  Since information is simply brokered by
the authority, the only right it has to legitimacy is by virtue of a
reputation. Thus expectation 1. above is based, in general, on
the rumours that an agent has heard.

Most of the trust is from users to the authority, thus there is a
clear subordination of agents in this model. This is the nature or
centralization.

\subsection{Web of Trust}

\begin{figure}[ht]
\begin{center}
\psfig{file=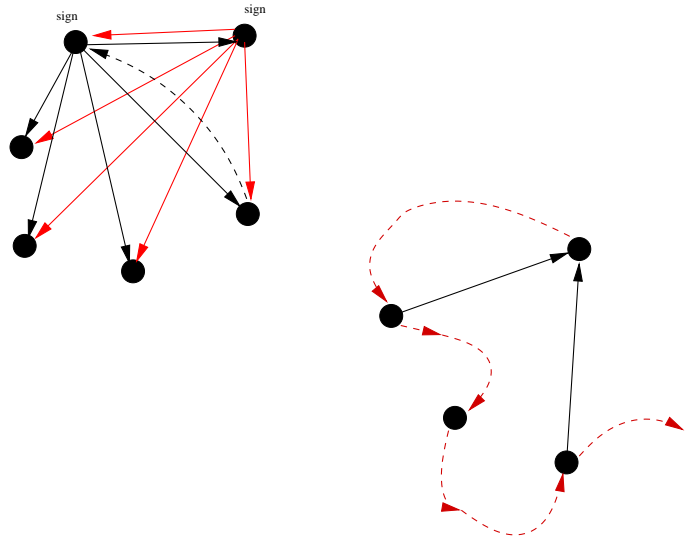,width=9cm}
\caption{\small In a web of trust an agent signals a
promise to all other agents that it has trusted the authenticity of
the originator's identity. As a key is passed around (second figure)
agents can promise its authenticity by signing it or not.
\label{webtrust}}
\end{center}
\end{figure}

Scepticism in centralized solutions (distrust perhaps) led to the
invention of the epidemic trust model, known as the Web of Trust (see
fig. \ref{webtrust})\cite{abdul1}. In this model, each individual
agent is responsible for its own decisions about trust. Agents
confirm their belief in credentials by signing one another's
credentials. Hence if I trust $A$ and $A$ has signed $B$'s key
then I am more likely to trust $B$.

As a management approximation, users are asked to make a judgement
about a key from one of four categories: i) definitely trustworthy,
ii) somewhat trustworthy, iii) un-trustworthy, iv) don't know.

An agent then compares these received valuations to a threshold value
to decide whether or not a credential is trustworthy to it.

The promises are between the owner of the credential and a random agent:
\beq
{\rm Owner} &\promise{\rm Identity} &{\rm Agent} \\
{\rm Agent} &\promise{U({\rm Identity})}  &{\rm Owner} \\
{\rm Agent} &\promise{\rm Signature}  &{\rm Owner} \\
{\rm Owner} &\promise{U({\rm Signature})} &{\rm Agent}
\eeq
The owner must first promise its identity to an agent it meets.  The
agent must promise to believe and use this identity credential. The
agent then promises to support the credential by signing it, which
implies a promise (petition) to all subsequent agents. Finally, the
owner can promise to use the signature or reject it. Trust enters here
in the following ways:

\begin{enumerate}
\item The agent expects that the identity of the owner is correct and trusts it.
This leads to a Use promise.
\item The Owner expects that the promise of support is legitimate and trusts it.
This leads to a Use promise.
\end{enumerate}
What is interesting about this model is that it is much more
symmetrical than the centralized scheme.  It has certain qualities
that remind us of our definition of global trust in section \ref{central}.
However, it is not equivalent to our model, since the very nature of
the web of trust is dictated by the transactions in the model, which
are automatically bilateral (ours need not be). Moreover, the
information is passed on in a peer to peer way, where as our global
idealization makes trust valuations common knowledge (global
reputations). In some respects, the web of trust is a pragmatic
approximation to the idealized notion of trust in section
\ref{central}. The main differences are:
\begin{itemize}
\item In the Web of trust, a limited number of expectation values is allowed
and the user does not control these, i.e. there are few policy choices 
for agent expectation allowed.

\item An agent does not see a complete trust or promise graph. It sees only the local
cluster to which it is connected. This is sufficient to compute a global
trust for that component of the graph.

\item The Web of Trust graph is always bilateral, with arrows moving in both directions,
thus no one is untrusted, or un-trusting.

\item The information to construct a fully self-consistent measure of trust
is not available in the system.  Hence there is no clear measure of
who is more trustworthy in the web of trust.

\end{itemize}

Some of these limitations could no doubt be removed.  A Bayesian
approach could naturally lead to a better approximation.  However, a
basic flaw in these implementation mechanisms is the need to trust of
the mediating software itself. Since, as we have shown, trust is not
necessarily transitive, one ends up in most cases trusting the
software that is supposed to implement the trust management rather
than the parties themselves.

\section{Conclusions}

The concept of promises provides a foundation that has been unclear in
discussions of trust. It allows us to decouple the probabilistic
aspect from the network aspect of policy relationships, without
introducing instantaneous events. It provides (we claim) a natural
language for specific policies, extended over time. Promises have
types and denote information flow which in turn allows us to discuss
what is trusted and by whom. We believe the use of promises to be
superor to a definition based on actions, since the localization of
actions as space-time events makes trust ill-defined if the action has
either not yet been executed or after it has been executed.

Promises allow us to relate trust and trust-reputation in a generic
way, and suggest an algorithm from which to derive global network
properties, based on social network theory. This is a significant
improvement over previous models. Reputation is not uniquely coupled
to trust, of course -- it can be related to many different valuations
of promised behaviour, including wealth, kindness etc.

We show how bundles of promises can be combined using the rules for
probabilistic events (similar to fault tree analysis) and we model the
two main trust architectures easily. The PGP Web of Trust as well as
the Trusted Third Party can be explained as a special case the global
trust models derived here; however standard tools do not permit users
to see the entire web, or measure relative trust-worthiness in a
community using these implementations.

In future work there is the possibility to use this notion of trust in
explicit systems. The Unix configuration system cfengine\cite{cfwww}
uses the notion of promises and agent autonomy to implement a policy
based management system. The trustworthiness of hosts with respect to
certain different behaviours can be measured directly by neighbouring
agents to whom promises are made. More generally, if one has a
monitoring system that one believes trustworthy to begin with, it is
possible to observe whether an agent stops keeping its own promises
about security issues. This might be a signal to reevaluate one's
expectation that the system is trustworthy. These tests have been
partially imeplemented in cfengine and are presently being tested.

Trust is merely an expression of policy and it is therefore
fundamentally {\em ad hoc}. Promises reveal the underlying motives for
trust and whether they are rationally or irrationally formed.

{\bf Acknowledgement.}
We are grateful to J\"urgen Sch\"onw\"alder for his hospitality at the
International University of Bremen in the summer of 2006, 
where most of this work was done.
This work is supported in part by the EC IST-EMANICS Network of
Excellence (\#26854).


\bibliographystyle{unsrt}
\bibliography{LISAbib}

\begin{thebibliography}{10}

\bibitem{promiseidea}
J.~Bergstra and M.~Burgess.
\newblock A static theory of promises.
\newblock Technical report, arXiv:0810.3294v1, 2008.

\bibitem{burgessDSOM2005}
Mark Burgess.
\newblock An approach to understanding policy based on autonomy and voluntary
  cooperation.
\newblock In {\em IFIP/IEEE 16th international workshop on distributed systems
  operations and management (DSOM), in LNCS 3775}, pages 97--108, 2005.

\bibitem{siri1}
M.~Burgess and S.~Fagernes.
\newblock Pervasive computing management: A model of network policy with local
  autonomy.
\newblock {\em IEEE Transactions on Software Engineering}, page (submitted).

\bibitem{siri2}
M.~Burgess and S.~Fagernes.
\newblock Voluntary economic cooperation in policy based management.
\newblock {\em IEEE Transactions on Network and Service Management}, page
  (submitted).

\bibitem{siri3}
M.~Burgess and S.~Fagernes.
\newblock Autonomic pervasive computing: A smart mall scenario using promise
  theory.
\newblock {\em Proceedings of the 1st IEEE International Workshop on Modelling
  Autonomic Communications Environments (MACE); Multicon verlag 2006. ISBN
  3-930736-05-5}, pages 133--160, 2006.

\bibitem{lapadula1}
L.~LaPadula.
\newblock A rule-set approach to formal modelling of a trusted computer system.
\newblock {\em Computing systems (University of California Press: Berkeley,
  CA)}, {\bf 7}:113, 1994.

\bibitem{mcilroy1}
M.D. McIlroy.
\newblock Virology 101.
\newblock {\em Computing systems (University of California Press: Berkeley,
  CA)}, {\bf 2}:173, 1989.

\bibitem{winkler2}
I.S. Winkler.
\newblock The non-technical threat to computing systems.
\newblock {\em Computing systems (MIT Press: Cambridge MA)}, {\bf 9}:3, 1996.

\bibitem{patton04technologies}
M.~Patton and A.~J{\o}sang.
\newblock Technologies for trust in electronic commerce.
\newblock {\em Electronic Commerce Research Journal}, 4:9--21, 2004.

\bibitem{sang-can}
Audun J{\o}sang, Claudia Keser, and Theo Dimitrakos.
\newblock Can we manage trust?
\newblock In {\em Proceedings of the Third International Conference on Trust
  Management (iTrust), Versailes}, 2005.

\bibitem{huynh2004a}
T.~Dong Huynh, Nicholas~R. Jennings, and Nigel~R. Shadbolt.
\newblock Developing an integrated trust and reputation model for open
  multi-agent systems.
\newblock In Rino Falcone, Suzanne Barber, Jordi Sabater, and Munindar Singh,
  editors, {\em AAMAS-04 Workshop on Trust in Agent Societies}, 2004.

\bibitem{klwer05trustworthiness}
J.~Kl{\"u}wer and A.~Waaler.
\newblock Trustworthiness by default, 2005.

\bibitem{relativetrust}
J.~Kl{\"u}wer and A.~Waaler.
\newblock Relative trustworthiness.
\newblock In {\em Formal Aspects in Security and Trust: Third International
  Workshop, FAST 2005, Newcastle upon Tyne, UK, July 18-19, 2005, Revised
  Selected Papers, Springer Lecture Notes in Computer Science 3866}, pages
  158--170, 2006.

\bibitem{beth1}
T.~Beth, M.~Borcherding, and B.~Klein.
\newblock Valuation of trust in open networks.
\newblock In {\em Proceedings of the European Symposium on Research in Computer
  Security (ESORICS), LNCS}, volume 875, pages 3--18. Springer, 1994.

\bibitem{jossang1}
Audun J{\o}sang and Simon Pope.
\newblock Semantic constraints for trust transitivity.
\newblock In {\em APCCM '05: Proceedings of the 2nd Asia-Pacific conference on
  Conceptual modelling}, pages 59--68, Darlinghurst, Australia, Australia,
  2005. Australian Computer Society, Inc.

\bibitem{trust1}
D.~Fahrenholtz and A.~Bartelt.
\newblock Towards a sociological view of trust in computer science.
\newblock In {\em Proceedings of the Eighth Research Symposium on Emerging
  Electronic Markets (RSEEM 01)}, page~10, 2001.

\bibitem{axelrod2}
R.~Axelrod.
\newblock {\em The Evolution of Co-operation}.
\newblock Penguin Books, 1990 (1984).

\bibitem{hoyland1}
A.~H{\o}yland and M.~Rausand.
\newblock {\em System Reliability Theory: Models and Statistical Methods}.
\newblock J. Wiley \& Sons, New York, 1994.

\bibitem{burgessbook2}
M.~Burgess.
\newblock {\em Analytical Network and System Administration --- Managing
  Human-Computer Systems}.
\newblock J. Wiley \& Sons, Chichester, 2004.

\bibitem{schelling1}
{\em The Strategy of Conflict}.
\newblock Harvard Univesity Press, Cambridge, Mass., 1960.

\bibitem{burgessC12}
M.~Burgess, G.~Canright, and K.~Eng{\o}.
\newblock A graph theoretical model of computer security: from file access to
  social engineering.
\newblock {\em International Journal of Information Security}, 3:70--85, 2004.

\bibitem{bonacich1}
P.~Bonacich.
\newblock Power and centrality: a family of measures.
\newblock {\em American Journal of Sociology}, 92:1170--1182, 1987.

\bibitem{roles}
G.~Canright and K.~Eng{\o}-Monsen.
\newblock A natural definition of clusters and roles in undirected graphs.
\newblock {\em Science of Computer Programming}, 53:195, 2004.

\bibitem{burgessroles}
M.~Burgess, G.~Canright, and K.~Eng{\o}.
\newblock Inportance-ranking functions from the eigenvectors of directed
  graphs.
\newblock {\em Journal of the ACM (Submitted)}, 2004.

\bibitem{morgenstern1}
J.V. Neumann and O.~Morgenstern.
\newblock {\em Theory of games and economic behaviour}.
\newblock Princeton University Press, Princeton, 1944.

\bibitem{rapoport1}
A.~Rapoport.
\newblock {\em N-Person Game Theory: Concepts and Applications}.
\newblock Dover, New York, 1970.

\bibitem{itut1}
ITU-T.
\newblock {\em Open Systems Interconnection - The Directory: Overview of
  Concepts, models and service. Recommendation X.500.}
\newblock International Telecommunications Union, Geneva, 1993.

\bibitem{x509}
ITU-T Recommendation.
\newblock X.509 (1997 e): Information technology - open systems interconnection
  - the directory: Authentication framework.
\newblock Technical report, 1997.

\bibitem{rfc3280}
R.~Housley, W.~Polk, W.~Ford, and D.~Solo.
\newblock Internet x.509 public key infrastructure: Certificate and certificate
  revocation list (crl) profile.
\newblock http://tools.ietf.org/html/rfc3280, 2002.

\bibitem{abdul1}
A.~Abdul-Rahman.
\newblock The pgp trust model.
\newblock {\em EDI-Forum: the Journal of Electronic Commerce}, 1997.

\bibitem{cfwww}
M.~Burgess.
\newblock Cfengine www site.
\newblock {\em http://www.cfengine.org}, 1993.

\bibitem{grimmett1}
G.R. Grimmett and D.R. Stirzaker.
\newblock {\em Probability and random processes (3rd edition)}.
\newblock Oxford scientific publications, Oxford, 2001.

\bibitem{pearl2}
J.~Pearl.
\newblock {\em Probabilistic Reasoning in Intelligent Systems: Networks of
  Plausible Inference}.
\newblock Morgen Kaufmann, San Francisco, 1988.

\end{thebibliography}


\appendix

\section{Expectation}

Here we expand on our notion of an expectation function for completeness.
These details are not essential to the arguments in the rest of the paper.

An expectation function is a statistical concept that relies either on
a body of evidence, or alternatively on a belief informed by limited
observation. Such evidences or beliefs are summarized by a probability
distribution over the different possible outcomes. We shall consider
mainly the outcomes ``promise kept'' and ``promise not kept'', though
varying degrees are possible.

The notion of an expectation value is well known from the theory of
probability and can be based on either classical
frequentist-probability or Bayesian
belief-probability\cite{grimmett1}. There is, for this reason, no
unique expectation operator.

Why dabble in intangibles such as beliefs?  Computer systems are
frequently asked to trust one another without ever having met 
(for example when they automatically download patches and updates from
their operating system provider, virus update from third-parties
or even accept the word of trusted third parties in identification) 
-- thus
they have little or no empirical evidence to go on. Each time they
interact however, they are able to revise their initial estimates on
the basis of experience. In this regard, a Bayesian view of
probability is a natural interpretation, see e.g. \cite{pearl2}. This
is a subjective view of probability that works well with our subjective
agents.

\begin{defin}[Expectation function $E(X)$]
Given random variables
$X,Y$, an expectation operator or function has the properties:
\begin{enumerate}
\item If $X \ge 0$, $E(X) \ge 0$.
\item If $a,b \in \field{R}$, then $E(aX+bY)= aE(X)+bE(Y)$.
\item E(1) = 1.
\end{enumerate}
For a probability distribution over discrete classes $c = 1\ldots C$, it is the convex
sum
\beq
E(X) = \sum_{c=1}^C  p_c X_x ~~\Big | ~~~~\sum_{c=1}^C p_c = 1.
\eeq
\end{defin}
The expectation value of a Bernoulli variable (with value 0
or 1) is clearly just equal to the probability of obtaining 1 $p_1 =
Pr(X = 1)$.  In general, a promise might lead to more than one
outcome, several of which might be acceptable ways of keeping the
promise, however this possibility only complicates the story
for now, hence we choose to consider only the simplest
case of
\begin{defin}[Agent expectation $E_A(a \promise{b} c)$]
The agent expectation $E_A(X)$ is defined to be the agent $A$'s estimation
of the probability that a promise $a \promise{b} c$ will be kept.
\end{defin}
This can be realized in any number of different ways, e.g. as a
mapping from an ensemble of evidence of size $N$ (with
the binary outcomes 0 and 1) into the open interval:
\beq
E_A : \{0,1\}^N \rightarrow [0,1]
\eeq
or it could could be an ad hoc value selected from a table of
pre-decided values.

\subsection{Ensembles and samples of evidence}\label{ensemble}

An ensemble is a collection of experiments that test the value of a
random variable. In one experiment, we might evaluate the agent
expectation to be $p_1$. In another, we might evaluate it to be
$P_1$. What then is the probability we should understand from the
ensemble of both?  We expect that the appropriate answer is an average
of these two values, but what if we attach more importance to one
value than to the other? Probabilities discard an essential piece of
information: the size of the body of evidence on which they are
based. Let us consider this important point for a moment.

\subsubsection{Frequentistic interpretation}

In the frequentist interpretation of probability, all estimates are
based on past hard evidence. Probabilities are considered reliable as
estimators of future behaviour if they are based on a sufficiently
large body of evidence. Let lower-case $p_1 = n_1/n$, the probability of
keeping a promise, be based on a total of $n$ measurements, of which the {\em frequencies}
$n_1$ were positive and $n_0$ were negative, with $n_1+n_0=n$. Also,
let upper-case $P_1 = N_1/N$ be an analogous set of measurements for which
$N \not= n$. How should we now combine these two independent trials
into a single value for their ensemble?

In the frequentist interpretation of probability, the answer is clear:
we simply combine all the original data into one trial and see what
this means for the probabilities. Rationally, the combined probability
$E$ for the ensemble must end up having the value $E =
(n_1+N_1)/(n+N)$. If we express this result in terms of the probabilities,
rather than the frequencies, we have
\beq
E = \left(\frac{n}{n+N}\right) \; p_1 + \left(\frac{N}{n+N}\right)\; P_1 = \frac{n_1+N_1}{n+N}
\eeq
This leads us to the intuitive conclusion that the probabilities should
be combined according to a weighted average, in which the weights are
chosen to attach proportionally greater importance to the larger trial:
\beq
E = \alpha_1 p_1 + \alpha_2  P_1, ~~ ~~ \alpha_1+\alpha_2 = 1.
\eeq
In general, then, with $T$ trials of different sizes, the result would
be a convex combination of the expectations from each trial:
\beq
E = \sum_{i=1}^T \; \alpha_i p_i, ~~  ~~ \sum_{i=1}^T \alpha_i = 1. \label{itup}
\eeq
In the case that there are more possible outcomes than simply
0 and 1, the same argument applies for each outcome.

The problem occurs when we do not have complete knowledge of the
sample sizes $n,N,\ldots$ etc., for, in this case, we can only guess
the relative importances $\alpha_i$, and choose them as a matter of
policy. If, for example, we could choose to make all the $\alpha_i$
equally important, in which case we have no control over the
importance of the expectations.

This so-called {\em frequentist} interpretation of expectation or
probability generally requires a significant body of evidence in the
form of independent events to generate a plausible estimate. However,
in most ad hoc encounters, we do not have such a body of
evidence. Trust is usually based on just a handful of encounters, and
one's opinion of the current evidence is biased by prior
expectations. Hence, we turn to the alternative interpretation or
Bayesian probability.

\subsubsection{Bayesian interpretation}

The policy formula in eqn. (\ref{itup}) is essentially a Bayesian
belief formula, which can be derived from the classic Bayes
interpretation for {\em a posteriori} belief.

Suppose we devise an experimental test $e$ to determine whether a
hypothesis $H$ of expected trustworthiness is true. We repeat this
test, or borrow other agent's observations, thus collecting $n$ of
these $e_1\ldots e_n$. The result for $P(H|e_n,e)$, our belief in the
trustworthiness-hypothesis given the available evidence, changes by
iteration according to:

\beq
P(H|e_n,e) = \frac{P(H|e_n) \times P(e|e_n,H)}
{P(e|e_n,H)P(H|e_n)+P(e|e_n,\neg H)P(\neg H|e_n)}\label{update}
\eeq
where we 
feed back one value $P(H|e_{n-1},e)$ from the previous iteration as $P(H|e_n)$,
and we
must revise potentially two estimates on each iteration:
\begin{itemize}
\item $P(e|e_n,H)$ is our estimate that the test $e$ will show positive as a direct
result of the Hypothesis being true, i.e. because the host was
trustworthy.

\item $P(e|e_n,\neg H)$ is our estimate of how often $e$ is true due to other
causes than the hypothesis $H$ of trustworthiness, e.g. due to
trickery.

\end{itemize}
Note that $P(\neg H|e_n) = 1 -P( H|e_n)$. This gives us a definite
iterative procedure based on well-accepted Bayesian belief networks
for updating our policy on trust\cite{pearl2}.  It can easily be seen
that eqn. (\ref{itup}) has this form, but lacks a methodology for
rational policy-making.

The advantage of a Bayesian interpretation of policy then, is that
it fits well with the notion of trust as a policy decision.

\subsection{Policy and rationality}

What kind of policy should be employed in defining the expectation of
future behaviour? Probability theory is built on the assumption that
past evidence can motivate a prediction of the future. At the heart of
this is an assumption that the world is basically constant.  However,
future prediction is the essence of gambling: there are scenarios in
which evidence of the past is not an adequate guide to future
behaviour. An agent might also look elsewhere for guidance.

\begin{itemize}
\item {\em Initialization}: An agent of which we 
have initially no experience might be assigned an initial trust value
of $1, \2,$ or $0$ if we are respectively trusting, neutral or
un-trusting by nature.

\item {\em Experience}: One's own direct experience of 
a service or promise has primacy as a basis for trusting an agent in a
network.  However, an optimistic agent might choose not to allow the
past to rule the future, believing that agents can change their
behaviour,  e.g. ``the agent was having a bad day''.

\item {\em Advice}: An agent might feel that it is not the best judge
and seek the advice of a reputable or trustworthy agent. ``Let's see
what X thinks''. We shall use this idea in section \ref{central}
to define a global trustworthiness.

\item {\em Reputation}: 
Someone else's experience with a promise can serve as an initial value
for our own trust.

\item {\em Damnation}: Some agents believe that, 
if an agent fails even once to fulfil a promise, then it is completely
un-trustworthy. This extreme policy seems excessive, since there might
be reasons beyond the control of the agent that prevent it from
delivering on its promise.

\end{itemize}

If we lack any evidence at all about the trustworthiness of an agent
with respect to a given promise, we might adopt a policy of
using the agent's record of keeping other kinds of promises.

\begin{proposal}[Transference of evidence]
In the absence of direct evidence of type $t(b)$, in a promise body
$b$, one may use a policy determined mixture of values from other
types as an initial estimate.
\end{proposal}
The rationality of such a procedure can easily be questioned,
but there is no way to rule out the ad hoc decision as a matter
of policy.

\end{document}